\documentstyle[aps,multicol,prl]{revtex}
\begin{document}
\title
{
Morphology and scaling in the noisy Burgers equation\\
Soliton approach to the strong coupling fixed point
}
\author{Hans C. Fogedby}
\address{
\thanks{Permanent address}
Institute of Physics and Astronomy,
University of Aarhus, DK-8000, Aarhus C, Denmark\\
and\\
NORDITA, Blegdamsvej 17, DK-2100, Copenhagen {\O}, Denmark
}
\date{\today}
\maketitle
\begin{abstract}
The morphology and scaling properties of the noisy Burgers equation
in one dimension are treated by means of a nonlinear soliton approach based
on the Martin-Siggia-Rose technique. In a canonical
formulation the strong coupling fixed point is accessed by means of
{\em a principle of least action} in the asymptotic nonperturbative
weak noise limit. The strong coupling scaling behaviour and the growth
morphology are described by a gas of nonlinear
soliton modes with a gapless dispersion law and a superposed
gas of linear diffusive modes with a gap. The dynamic
exponent is determined by the gapless soliton dispersion law, whereas
the roughness exponent and a heuristic expression for the scaling function
are given by the form factor in a spectral representation of the
interface slope correlation function. The scaling function has the
form of a L\'{e}vy flight distribution.
\end{abstract}
\draft
\begin{multicols}{2}
There is a current interest in the scaling properties and general morphology
of nonequlibrium models. Here the noisy Burgers equation
in one dimension provides maybe the simplest continuum description of
an open driven nonlinear system \cite{burgers,krug,forster,kpz,fogedby1}.
This equation has the form of a conserved Langevin equation
\begin{equation}
\partial u/\partial t = \nu\nabla^2u + \lambda u\nabla u +\nabla\eta ~ ,
\label{burgers}
\end{equation}
where $\nu$ is a damping constant characterizing the linear diffusive term,
$\lambda$ a coupling strength for the nonlinear mode coupling term,
and $\eta$ a Gaussian
white noise driving the system into a stationary state and correlated 
according to
\begin{equation}
\langle\eta(xt)\eta(x't')\rangle=\Delta\delta(x-x')\delta(t-t') ~ .
\label{noise}
\end{equation}
The equation (\ref{burgers}) has been discussed extensively in particular as a
model for the stochastic dynamics of the slope field $u=\nabla h$ for
the self-affine growth of an interface subject to annealed noise from
the drive or environment described by the Kardar-Parisi-Zhang (KPZ) equation
\begin{equation}
\partial h/\partial t=\nu\nabla^2h+(1/2)\lambda(\nabla h)^2+\eta ~ .
\label{kpz}
\end{equation}

The main emphasis has been on the scaling properties of the slope field
in the large $x$ - long $t$ limit as embodied in the dynamical scaling
form \cite{forster,kpz}
\begin{equation}
\langle u(xt)u(x't')\rangle=|x-x'|^{-2(1-\zeta)}f(|t-t'|/|x-x'|^z) ~  ,
\label{scaling}
\end{equation}
where $\zeta$ is the roughness exponent, $z$ the dynamic exponent,
and $f$ the scaling function; $f(w)\propto w^{-2(1-\zeta)/z}$ for
large values of its argument. In this context the nonlinear Galilean
symmetry
of (\ref{burgers}) 
\begin{equation}
x\rightarrow x-\lambda u_0t ~ ,
~~~~~~~
u\rightarrow u+u_0
\label{galilean}
\end{equation}
with $\lambda$ entering as a {\em structural constant} implies the
scaling law
\begin{equation}
\zeta + z = 2 ~ .
\label{scaling-law}
\end{equation}
Another property specific to one dimension is the 
stationary probability
distribution \cite{krug,kpz,fogedby1}, 
\begin{equation}
P(u)\propto\exp{[-(\nu/\Delta)\int dx u^2 ~]} ~ .
\label{distribution}
\end{equation}
{\em independent} of $\lambda$.
Hence $u$ is an {\em independent} Gaussian variable,
$\langle u(x)u(x')\rangle=(\Delta/2\nu)\delta(x-x')$,
and the height
field $h=\int dx u$ performs {\em random walk} with
exponent $\zeta = 1/2$. The scaling law 
(\ref{scaling-law}) then yields $z=3/2$,
defining the Burgers-KPZ universality class. 
This result also follows from dynamic renormalization group calculations
\cite{forster,kpz} but the associated infrared stable
{\em strong coupling fixed point}
does not seem accessible by the methods of renormalized perturbation
theory and the $\epsilon$-expansion.

In the linear
Edwards-Wilkinson (EW) case for $\lambda=0$ the slope correlations are
\begin{equation}
\langle u(k\omega)u(-k-\omega)\rangle = 
\frac{\Delta k^2}{\omega^2+(\nu k^2)^2} ~ ,
\label{EW-correlation}
\end{equation}
yielding the exponents $\zeta = 1/2$ and $z = 2$,  characterizing
the EW universality class with scaling function
\begin{equation}
f(w)=(\Delta/2\nu)(4\pi\nu)^{-1/2}w^{-1/2}\exp{[-1/4\nu w]} ~ .
\label{EW-scaling}
\end{equation}

In a recent letter \cite{fogedby1} we approached the strong
coupling fixed point behaviour
by means of a mapping  of the Burgers equation (\ref{burgers}) onto a
solid-on-solid model and further onto a discrete spin $1/2$ chain model.
In a harmonic oscillator representation valid for large spin combined with
a continuum limit this approach eventually yields a Hamiltonian
description and a set of coupled equations of motion with soliton
solutions with dispersion $E\propto p^{3/2}$, yielding the exponent
$z=3/2$, characterizing the strong coupling fixed point. The approach
furthermore gave a tentative picture of a growing interface in terms of a
soliton gas of paired solitons representing a growing step in the height
field.

Here we present a {\em unified} approach to the noisy Burgers
equation based on a Martin-Siggia-Rose path integral. 
This method provides a generalization of the stationary distribution
(\ref{distribution}) to the time-dependent nonlinear case and 
yields the following physical picture and scaling results:\\
1) In the weak noise limit the general morphology of a growing 
interface in the stationary
regime is described by a Landau-type quasi-particle many-body
theory in terms
of a gas of nonlinear soliton modes with superposed linear diffusive
modes.\\
2) The noise-driven fluctuations correspond 
to ``quantum fluctuations'' in the underlying non-Hermitian
relaxational ``quantum description'', yielding the quasi-particle picture.\\
3) The scaling exponents and the scaling function follow as a by-product
from the soliton dispersion law and the spectral representation
of the correlations. The many-body formulation explains the robustness
of the roughness exponent under a change of universality class.

From a field theoretical point of view we identify the noise strength
$\Delta$  in (\ref{noise}) as the effective {\em small parameter}. Furthermore,
the fundamental probability distribution or path integral 
has an {\em essential singularity}
for $\Delta = 0$.
Hence our approach is based on a {\em nonperturbative
saddle point or steepest descent approximation} to the path integral.
It is precisely in this respect that the dynamic renormalization group
method based on an expansion in $\lambda$ {\em and} in the noise contraction
$\Delta$ fails to access the strong coupling fixed point.

The special role of the noise is easy to understand. The {\em turning on}
of the noise in the Burgers equation is a {\em singular process}
in the sense that even {\em weak noise} $\Delta\sim 0$ will eventually
drive the system into a {\em stationary state} whereas for $\Delta =0$
the noiseless deterministic Burgers equation (assuming vanishing currents
at the boundaries) describes a damped interface, although with
{\em pattern formation} or {\em dissipative structures} formed by
the nonlinear {\em cascade} term in the transient time regime.
In the presence of noise the system passes from the {\em transient}
to the {\em stationary} regime at a crossover time $t_{co}$
determined by the noise strength.
In the linear EW case $t_{co}$ is of order
$1/(\nu k^2)\log{(1/\Delta)}$ showing that $t_{co}\rightarrow\infty$
for $\Delta\rightarrow 0$; $k$ is the wave number of the particular
mode considered (note that $t_{co}\rightarrow\infty$ also
for $k\rightarrow 0$, characteristic of a  conserved hydrodynamical mode).
In the linear EW case the singular character of the noise strength
$\Delta$ is reflected in the stationary distribution (\ref{distribution})
which has an {\em essential singularity} for $\Delta = 0$.

It is well-known that the {\em noiseless}  Burgers equation
for $\Delta = 0$
\cite{burgers,kpz,fogedby2}, 
$
\partial u/\partial t = \nu\nabla^2u + \lambda u\nabla u 
$,
supports parity-breaking {\em right hand} solitons
connected by ramp solutions, corresponding to cusps 
connected by convex parabolic segments in the height field.
Superposed on the solitons are 
linear diffusive modes with a gap in the spectrum.
This configuration constitutes the growth pattern in the
transient time regime.
For $\lambda = 0$ the solitons disappear and the equation
reduces to the linear diffusion equations supporting gapless
diffusive modes.

In the stationary noise-driven case for
$\Delta\neq 0$ the soliton excitations  and 
diffusive modes also play an important role in the analysis of the
physics and scaling behaviour of the strong coupling fixed. In the noisy
case the solitons appear as stationary or saddle point solutions
to the path integral in the weak noise limit; the diffusive modes
correspond to Gaussian fluctuations about the stationary points.
Also the noise fluctuations lift the broken parity symmetry and excites 
both {\em right} and {\em left hand} solitons.
As regards the diffusive modes both decaying  and growing modes are excited
in accordance with the time reversal invariance of the stationary regime
as for example reflected in the evenness in $\omega$ of the
slope correlations (\ref{EW-correlation}) in the linear case.

In terms of the Martin-Siggia-Rose path integral \cite{MSR} the 
slope correlations are given by \cite{footnote1}
\begin{equation}
\langle u(xt)u(x't')\rangle =
\frac{\int\prod_{xt}dud\varphi\exp{[i(\nu/\Delta)S]}u(xt)u(x't')}
{\int\prod_{xt}dud\varphi\exp{[i(\nu/\Delta)S]}}
\label{MSR}
\end{equation}
which provides the generalization of the stationary EW case
$
\langle u(x)u(x')\rangle =
\int\prod_xduP(u)u(x)u(x')
$
with $P(u)$ given by (\ref{distribution}).
The action $S$ in (\ref{MSR}) has the form
\begin{equation}
S = \int dxdt[ud\varphi/dt - {\cal H}(u,\varphi)]
\label{action}
\end{equation}
with Hamiltonian density
\begin{equation}
{\cal H} = -i(\nu/2)[(\nabla u)^2+(\nabla\varphi)^2] + 
(\lambda/2)u^2\nabla\varphi ~ .
\label{hamiltonian}
\end{equation}
The correspondence of (\ref{MSR},\ref{action},\ref{hamiltonian})
with 
the Feynman phase space path integral
\cite{das} allows us to consider $\Delta/\nu$ as an effective
{\em Planck constant}  and the slope field $u$ and  noise field
$\varphi$ as canonically conjugate variables satisfying the Poisson bracket
$\{u(x),\varphi(x')\}=\delta(x-x')$. It is also characteristic of a stochastic
growth problem that the Hamiltonian, being the generator in the relaxational
master equation, in general is complex. 
The imaginary harmonic
part drives the diffusive modes whereas the real nonlinear term describes
the intrinsic growth of the interface. The path integral replacing the
noisy Burgers equation is deterministic, the noise-driven fluctuations
being replaced by the contributing paths or configurations in the
path integral.

The canonical form of the path integral allows for 
{\em a principle of least action}. In other words, in the weak noise
limit $\Delta\rightarrow 0$ we can determine
``classical'' solutions or orbits from the stationary points of
the action. The weak noise solutions are then given by the
equations of motion, $du/dt = \{{\cal H},u\} = -\delta{\cal H}/\delta\varphi$
and $d\varphi/dt=\{{\cal H},\varphi\} = \delta{\cal H}/\delta u$,
\begin{equation}
\frac{\partial u}{\partial t} =-i\nu\nabla^2\varphi + \lambda u\nabla u ~ ,
~~
\frac{\partial \varphi}{\partial t} =+i\nu\nabla^2u + \lambda u\nabla\varphi ~ .
\label{field}
\end{equation}
The deterministic nonlinear coupled field equations 
(\ref{field}) for
the slope and noise fields  thus replace the noisy Burgers equation
(\ref{burgers}) in the asymptotic nonperturbative weak noise limit.
We note that i) the equations are invariant under a shift of $\varphi$
reflecting the conserved noise in (\ref{burgers}) and ii) like the noisy
Burgers equation,  invariant under the Galilean transformation 
(\ref{galilean}) supplemented with $\varphi\rightarrow\varphi$.

In addition to the constant solutions $(u,\varphi)=(u_0,\varphi_0)$,
corresponding to the zero-energy stationary state, the field equations
also support nonlinear soliton solutions of both parities.
The static solitons are given by
\begin{equation}
u=\pm u_+\tanh{(k_s(x-x_0))}~ ,
~~~~~~
k_s = \lambda u_+/2\nu ~ .
\label{soliton}
\end{equation}
Here $k_s$ is a wavenumber depending on the soliton amplitude, setting
the inverse length scale, and $x_0$ a center of mass position. 
Note that only the {\em right hand}
soliton appears in the noiseless case. Boosting the soliton to a 
finite velocity $v$
and denoting the boundary values by $u_\pm$ the general soliton
condition is \cite{fogedby1}
\begin{equation}
u_++u_- = -2v/\lambda ~ ,
\label{soliton-condition}
\end{equation}
relating the soliton amplitude and off-set to the propagation velocity.

For $\lambda = 0$ the field equations yield 
diffusive modes $u = A\exp{(-i\omega t+ikx)} + c.c.$
\label{linear}
with gapless dispersion
\begin{equation}
\omega = -i\nu k^2 ~ .
\label{dispersion}
\end{equation}
The Hamiltonian
(\ref{hamiltonian}) is harmonic, the path integral (\ref{MSR})
Gaussian,
and it is an easy task to evaluate the slope correlations
(\ref{EW-correlation}).

It is a well-known feature of field theoretical saddle point
soliton or instanton calculations that 
multi-soliton solutions also contribute to the stationary
points \cite{das}.
Here this implies that the extremal
path in the weak noise limit corresponds to a dilute soliton gas
of {\em right} and {\em left hand} solitons matched according to
the soliton condition (\ref{soliton-condition}). In other words,
the path integral formulation directly produces a physical
picture of the morphology of a growing interface. The natural variable
is the slope field $u$. A single soliton configuration
connecting two stationary slope configuration corresponds to the propagating
top or bottom of a growing step in the height field $h$. A
{\em single} moving step in $h$ is represented by a pair of solitons
matched according to (\ref{soliton-condition}). In a similar fashion
a moving plateau in $h$ corresponds to  four solitons, a growing tip
to a three-soliton configuration, etc. The morphology
of a growing interface can thus be interpreted as a dilute gas of
solitons corresponding to moving steps in the height field.
We note that the noise radically changes the morphology from the noiseless
transient case in that the noise excites both ``up'' and ``down'' cusps in $h$.
Superposed on the soliton gas are linear diffusive ``ripple'' modes
with a gap in the spectrum.

The canonical formulation of the path integral enables us
to associate energy, momentum, and action with a given soliton
configuration. This allows for a  {\em dynamical selection criterion}
similar to the Boltzmann factor $\exp{(-E/T)}$ in equilibrium statistical
mechanics which associates an energy $E$ with a given configuration 
contributing
to the partition function; in the dynamical case the action $S$ provides the
weight function for the dynamical configuration.

The energy is given by $E=\int dx{\cal H}$ whereas the momentum follows
from the Poisson bracket algebra, $P=\int dx u\nabla\varphi$. For a single
soliton we have in terms of the boundary values, 
$E=\pm i(\lambda/6)(u_+^3-u_-^3)$ and
$P=\pm i(1/2)(u_+^2-u_-^2)$.
In the case of a single moving soliton satisfying $u_- = 0$ 
we obtain in particular
\begin{equation}
E=(4i/3)|v|^3/\lambda^2~ ,
~~~~~
P=-(2i)v|v|/\lambda^2 ~ .
\label{eps}
\end{equation}
We note the nonlinear velocity dependence characteristic of soliton
solutions \cite{fogedby3,footnote2}.
Eliminating $v$ we obtain the soliton dispersion law
$E = \lambda(\sqrt{2}/3)\exp{(i\theta)}P^{3/2}$, where
the phase $\theta =\pi+(\pi/4)(v/|v|)$.

The linear diffusive mode spectrum in the presence of the solitons
is  analyzed by a linear stability analysis of the
field equations (\ref{field}). Like in the noiseless case
\cite{fogedby2} the associated eigenvalue problem is exactly soluble
\cite{fogedby4}. 
In addition to  zero-eigenvalue
bound states corresponding to the soliton translation modes
lifting the broken translational symmetry, the spectrum also
has a band of phase shifted diffusive modes with a gap in the dispersion
law
\begin{equation}
\omega = -i\nu(k^2 + k_s^2) ~ ,
\label{gap-dispersion}
\end{equation}
depending on the soliton amplitude according to (\ref{soliton}).

In order to complete the analysis of the noisy Burgers equation it remains
to discuss the contribution of the Gaussian fluctuations in the path integral
about the solitons. The basic information lies in the path integral and will
be discussed elsewhere. Here we take a heuristic point of view and discuss
the fluctuations in terms of the underlying non-Hermitian relaxational
``quantum field theory'' leading to the path integral by the usual Feynman
procedure \cite{das}. The canonical fields $u$ and $\varphi$
must now be interpreted as ``quantum operators'' satisfying a commutator 
algebra, the Hamiltonian (\ref{hamiltonian}) is a non-Hermitian
operator,
and the field equations (\ref{field}) are the associated
Heisenberg equations of motion. The ``classical'' localized soliton solution
(\ref{soliton}) becomes a delocalized ``quantum'' quasi-particle with
frequency $\Omega=E/(\Delta/\nu)$ and wavenumber $K=P/(\Delta/\nu)$.
For a pair of matched quasi-particles representing a growing step
the ``quantum'' dispersion law is
\begin{equation}
\Omega = \lambda(1/3)(\Delta/\nu)^{1/2}
e^{i\theta}K^\frac{3}{2} ~ .
\label{sol-dispersion}
\end{equation}
Evaluating the group velocity of a wave packet representing
the ``quantum soliton'' we find the classical propagation velocity,
in complete accordance with an effective {\em correspondence principle}.
In a similar way the linear diffusive modes are ``quantised'' becoming
quasi-particles corresponding to ``ripple'' modes on the solitons
with dispersion law (\ref{gap-dispersion}) \cite{fogedby3}. 

The picture of the growing interface is thus complete. The noise-induced 
fluctuations admits a ``quantum mechanical'' interpretation. The dominant 
morphology consists of a fluctuating ``quantum'' gas of solitons with
a gapless dispersion law with exponent $3/2$. Superposed on the soliton
gas are linear diffusive ``ripples'' modes with a gap in the dispersion
depending on the soliton amplitude. In the limit of vanishing growth
for $\lambda\rightarrow 0$ the solitons disappear and the diffusive modes 
become gapless corresponding to thermodynamic fluctuations governed by
the EW equation with Boltzmann weight (\ref{distribution}).

In the ``quantum interpretation'' the slope correlation function
$\langle u(xt)u(00)\rangle$ is given by a time-ordered Green's function
$\langle 0|T\hat{u}(xt)\hat{u}(00)|0\rangle$ \cite{das}. Here $\hat{u}$ is the
time-dependent ``quantum'' slope field and $|0\rangle$ denotes the
appropriate stationary zero-energy state. Assuming $t>0$, using the 
Hamiltonian and momentum
operators to ``displace'' the field at ($x,t$) to ($0,0$), and inserting
a complete set of  momentum ``quasi-particle'' states, we derive
an effective {\em spectral representation} which allows us to discuss the
scaling properties in simple terms,
\begin{equation}
\langle u(xt)u(00)\rangle =
\int dK G(K)e^{-i(\Omega t+Kx)} ~ .
\label{spectral}
\end{equation}
Here $G(K)$ is an effective {\em form factor} which in general is complex
owing to the non-Hermitian Hamiltonian (\ref{hamiltonian}).
At long distances (\ref{spectral}) samples the small wave number
region in the integral and assuming that the form factor is regular for
small $K$ and a general dispersion law of the form

\begin{equation}
\Omega=\tilde{\Delta}+K^\mu ~,
\label{general-dispersion}
\end{equation}
characterized by the gap $\tilde{\Delta}$ and the exponent $\mu$,
we obtain, absorbing $x$ in a rescaling of $K$
\begin{equation}
\langle u(x,t)u(0,0)\rangle \propto
e^{-i\tilde{\Delta}t}x^{-1}\int dK e^{-iK^\mu(t/x^\mu)-iK} ~ .
\label{final-spectral}
\end{equation}
First, in the presence of a gap there is no scaling due to the
exponential prefactor in (\ref{final-spectral}), consequently, 
only gapless excitations contribute to the scaling behaviour. For
$\tilde{\Delta}=0$ comparing (\ref{final-spectral}) with 
(\ref{scaling}) we infer $z=\mu$ and $\zeta = 1/2$.

In the linear EW case the diffusive gapless modes with
dispersion (\ref{dispersion}) exhaust the spectrum
and we obtain $\mu=z=2$, corresponding to the EW universality class.
The spectral form (\ref{final-spectral}) yields the scaling
function in (\ref{EW-scaling}).

In the nonlinear Burgers-KPZ case the nonlinear soliton modes 
with gapless dispersion (\ref{sol-dispersion}) exhaust the bottom
of the spectrum and yields $\mu=z=3/2$ for the Burgers-KPZ
universality class. The linear diffusive modes develop a
gap, become sub-dominant and do not contribute to the scaling.
This mechanism explains the change in universality class:
The universality class is determined by the dominant gapless
excitation. 
The spectral form (\ref{spectral}) also elucidates the {\em robustness} of the
exponent $\zeta$ which is the same for both universality classes.
For the stationary fluctuations we set $t=0$ in (\ref{spectral})
and the resulting scaling function yielding $\zeta$ does
{\em not} depend on the specific  quasi-particle
dispersion law.

Finally, we also obtain a heuristic expression for the scaling function
\begin{equation}
f(w) = \int dK e^{-i(K^zw+K)} 
\end{equation}
which incidentally has the same form as the probability distribution for 
L\'{e}vy flights \cite{fogedby5}.

In this letter we have presented the outline of a novel approach
to the growth morphology and scaling behaviour of the noisy Burgers
equation in one dimension; details will appear
elsewhere. Using the Martin-Siggia-Rose
technique in a canonical form we have demonstrated that the so far 
elusive strong coupling fixed point behaviour is associated with
an essential singularity in the noise strength and can be accessed by 
appropriate field theoretical soliton techniques. 

Discussions with J. Krug, M. Kosterlitz and A. Svane are gratefully 
acknowledged.

\end{multicols}

\begin{references}{}
\bibitem{footnote1}
In order to obtain the canonical form we have shifted the noise field as in
{\cite{fogedby1}}.
\bibitem{footnote2}
$E$ and $P$ can be made real by rotating the noise field,
$\varphi\rightarrow i\varphi$ in (\ref{MSR}).
\end{references}
\end{document}